\documentstyle[12pt]{article}
\newskip\humongous \humongous=0pt plus 1000pt minus 1000pt

\newif\ifdtup

\def\ee{e^+e^-}
\def\as{\alpha_S}

\def\lms{\Lambda_{\overline{\mbox{\scriptsize MS}}}}
\def\frac#1#2{ {{#1} \over {#2} }}

\def\nmax{\raisebox{-1.3ex}{\rlap{$\;\;\; \bf n \;\;$}} \raisebox{0ex}
{$\; {\rm Max} \;$}}
\def\t1{\raisebox{-1ex}{\rlap{\tiny $T \rightarrow 1$}} \raisebox{0ex}
{$\;\;\, \simeq \;\;$}}
\def\smally{\raisebox{-1.2ex}{\rlap{\tiny $\ycut \ll 1\;\;$}} \raisebox{0ex}
{$\;\;\;\;\; \simeq \;$}}
\def\ltap{\raisebox{-.4ex}{\rlap{$\,\sim\,$}} \raisebox{.4ex}{$\,<\,$}}
\def\gtap{\raisebox{-.4ex}{\rlap{$\,\sim\,$}} \raisebox{.4ex}{$\,>\,$}}
\def\ycut{y_{\rm cut}}
\def\VEV#1{\left\langle #1\right\rangle}
\def\naive{na\"{\i}ve}
\def\naively{na\"{\i}vely}
\def\half{\frac{1}{2}}
\def\cpc#1#2#3{Computer Phys.\ Comm.\ #1 (19#3) #2}

\def\np#1#2#3{Nucl.\ Phys.\ B#1 (19#3) #2}
\def\pl#1#2#3{Phys.\ Lett.\ #1B (19#3) #2}
\def\pr#1#2#3{Phys.\ Rev.\ D #1 (19#3) #2}
\def\prep#1#2#3{Phys.\ Rep.\ #1 (19#3) #2}
\def\prl#1#2#3{Phys.\ Rev.\ Lett.\ #1 (19#3) #2}

\def\zp#1#2#3{Zeit.\ Phys.\ C#1 (19#3) #2}
\def\jp#1#2#3{J.\ Phys.\ G#1 (19#3) #2}
\begin{document}
\begin{titlepage}
\renewcommand{\thefootnote}{\fnsymbol{footnote}}
\begin{flushright}
     DFF 211/10/94 \\
     October 1994
\end{flushright}
\vspace*{\fill}
%\par \vskip 10mm

\begin{center}
{\Large \bf \boldmath JET PHYSICS AT LEP AND SLC\footnote{Invited talk at the
18th Johns Hopkins Workshop on Current
Problems in Particle Theory, {\em Theory meets experiments}, Florence,
September 1994.}
}
\end{center}
\par \vskip 2mm
\begin{center}
{{\bf Stefano Catani}\\
\vskip 2mm
        {\it INFN, Sezione di Firenze} \\
        {\it and Dipartimento di Fisica, Universit\`{a} di Firenze,} \\
        {\it Largo E. Fermi} 2, I-50125 \it Florence, Italy
        \
        }
\end{center}
\par \vskip 2mm
\begin{center} {\large \bf Abstract} \end{center}
\begin{quote}
Experimental results on jet physics at LEP and SLC are reviewed and compared
with perturbative QCD predictions. The discussion includes determinations of
the strong coupling $\as$, measurements of event shape distributions and jet
cross sections, studies of subjet multiplicities and tests of QCD coherence.
\end{quote}
\vspace*{\fill}
\end{titlepage}
\pagestyle{plain}
\renewcommand{\thefootnote}{\fnsymbol{footnote}}
{\bf \noindent 1. Introduction}
\vskip 2.5mm
The large electron-positron colliders LEP and SLC have provided
many detailed studies of hadronic final states. These include jet cross
sections, direct-photon production, hadron spectra and, more
recently, studies of fragmentation functions and their scaling
violations. Most of these topics are discussed in several review
papers (see, for instance, Refs.~[\ref{Heb}-\ref{Webber}]). In this
contribution, rather than presenting a general review on hadronic physics
at LEP and SLC, I concentrate only on jet physics. The reason is
the following.

On the experimental side, LEP and SLC are remarkable machines for studying
hadronic interactions. They operate at a very high centre-of-mass energy,
${\sqrt s} \simeq 91$~GeV, and provide a very large event statistics (due to
the large cross section at the $Z^0$ resonance) and a very clean
environment (as in any $\ee$ collider, the final state is not
contaminated by the debris of initial-state hadrons).

On the theoretical side, if we consider jet physics (i.e. high-energy
hadronic observables which are infrared and collinear safe) we are
dealing with a very predictive theory, namely perturbative Quantum
Chromodyanamics (QCD). We have at our disposal a well defined theoretical
framework (with essentially only one free parameter, the strong
coupling $\as$) and quite accurate calculations.

Combining these experimental and theoretical facilities, we are thus in a
position to perform `precision tests' of QCD. The aim of the present
contribution is to show that at LEP/SLC we can investigate strong-interaction
physics to an accuracy better than $10\%$, which is quite an achievement
in this field.

The outline of this review is as follows. In Sect.~2, I recall the general
theoretical features of jet observables and their treatment within
perturbative QCD. Section 3 is devoted to the measurements of $\as$ from
fully inclusive quantities, namely the hadronic widths of the $Z^0$ boson and
the $\tau$ lepton. More detailed tests of QCD from event shape distributions
are considered in Sect.~4. Here, I present results based on analyses in
two-loop
order and including the all-order resummation of logarithmically enhanced
contributions. A summary of $\as$ determinations at LEP and SLC is also
included
in this Section.  Section 5 deals with the definition of jets and
studies of jet
cross sections. A discussion on subjet multiplicities, QCD coherence and
properties of quark and gluon jets is presented in Sect.~6. Finally, in
Sect.~7,
I briefly summarize the requirements for further progress in the field.

\vskip 4.0mm
{\bf \noindent 2. Jet observables in perturbative QCD}
\vskip 2.0mm

Jet quantities are, by definition, hadronic observables which turn out
to be infrared and collinear safe [\ref{SW}]. In other words, their actual
value does not vary if the final state changes by the addition of one
more particle which is soft or collinear to another particle.

The main theoretical feature of jet observables is that they can be computed
in QCD perturbation theory. This is not a trivial statement. In general,
evaluating QCD Feynman diagrams in terms of partons (quarks and gluons),
one finds integrals which are divergent in the low-momentum and/or small-angle
regions. In the case of jet quantities, the coherent sum over different
soft and collinear partonic states leads to the cancellation of these
divergences. As a result, jets observables are finite (calculable)
at the partonic level order by order in perturbation theory.

Denoting by $R$ a generic (dimensionless) jet variable, we can compute
its corresponding perturbative expansion in the form
\begin{equation}
\label{1}
R = 1 + \as(Q) \, R_1 + \as^2(Q) \, R_2 + \dots \;\;\;.
\end{equation}
The first non-trivial term $(\as R_1)$ in Eq.~(\ref{1}), the second term
$(\as^2 R_2)$, etc. represent respectively the leading-order (LO)
contribution to $R$, the next-to-leading order (NLO) contribution and so on.
The typical energy scale, at which the observable $R$ is considered,
is denoted by $Q$.

The expansion parameter in Eq.~(\ref{1}) is $\as(Q)$, the QCD running
coupling. It is the sole (apart from quark masses) free parameter
in the theory. QCD does not predicts its actual value but only its
energy behaviour. Indeed, due to the property of asymptotic freedom,
the running coupling $\as(Q)$ decreases as the energy $Q$ increases,
according to the approximate logarithmic behaviour $\as(Q) \sim
(\beta_0 \ln Q^2/\Lambda^2)^{-1}, \;\Lambda$ being the fundamental
scale of QCD.

Asymptotic freedom is due to the non-abelian gauge structure of the theory.
Thus, the experimental observation of the running of $\as(Q)$ according
to QCD is a fundamental test of the underlying gauge theory. For this reason,
in the past few years much effort has been devoted to measurements of
$\as(Q)$ (see Refs.~[\ref{Webber},\ref{BC}-\ref{Mont}] and Secs.~3 and 4).

The QCD running of $\as$ implies that the effective coupling $\as(Q)$ is
small at high energy $Q$. This property justifies the use of perturbation
theory for predicting jet observables, at least at asymptotic energies.
However, just because of its perturbative nature, the QCD running can
be hidden in higher-order corrections by the replacement
$\as(Q) = \as^{(0)} [1+K(Q)\, \as(Q) + \dots ]$, $\as^{(0)}$ being the value
of $\as$ at a fixed (and arbitrary) energy scale. It follows that a LO
calculation gives only the order of magnitude of a certain observable.
The accuracy of the perturbative QCD expansion is instead controlled
by the size of the higher-order corrections. Any definite QCD prediction
thus requires (at least) a NLO definition of $\as$ and a NLO
perturbative calculation.

In the following I shall consider the NLO definition of $\as$ as given in the
${\rm {\overline {MS}}}$ renormalization scheme. In this scheme the relation
between $\as(Q)$ and the QCD scale $\lms$ is [\ref{PDG}]
\begin{equation}
\label{2}
\as(Q) = \frac{1}{\beta_0 \ln Q^2/\lms^2} \left[
1 - \frac{\beta_1}{\beta_0^2}
\frac{\ln (\ln Q^2/\lms^2)}{\ln Q^2/\lms^2} +
{\cal O}\left(
\frac{\ln^2 (\ln Q^2/\lms^2)}{\ln^2 Q^2/\lms^2}\right)
\right] \;,
\end{equation}
where $12 \pi \beta_0 = 33-2N_f, \;24 \pi^2 \beta_1 = 153 - 19 N_f$ and
$N_f$ is the number of quarks whose mass is smaller than $Q$.

As for the power series expansion in $\as$ beyond the LO, one has to be more
precise than in Eq.~(\ref{1}). In general, the perturbative series has the
following structure
\begin{eqnarray}
\label{3}
R(\as(\mu), Q/\mu;y) &=& 1+\as(\mu) R_1(y)
%\nonumber \\ &\!&
+ \as^2(\mu) \left[ R_2(y)-
R_1(y) \,\beta_0 \ln Q^2/\mu^2 \right] \nonumber \\
&+& {\cal O}(\as^3(\mu) \ln^2 Q^2/\mu^2, \,\as^3(\mu)) \;\;,
%\nonumber
\end{eqnarray}
where $Q$ is the typical energy scale of the process and $y$ denotes any
other kinematic scale involved in the definition of the observable $R$.

In Eq.~(\ref{3}) we have made explicit the dependence on the
renormalization scale $\mu$ which has to be introduced in any higher-order
calculation in order to control (regularize) ultraviolet divergences.
Obviously, the observable $R$ is a renormalization group invariant quantity
if computed to all orders in $\as$. The $\mu$-dependence is an artifact of
the truncation of the perturbative expansion at a fixed order in $\as$. This
implies that $a)$ the renormalization scale dependence is formally an
higher-order effect (i.e. $dR/d\ln \mu={\cal O}(\as^3)$ in Eq.~(\ref{3}))
and, correspondingly, $b)$ higher-order terms have an explicit
$\ln (Q/\mu)$-dependence
(see the ${\cal O}(\as^3 \ln^2 Q^2/\mu^2)$ term in Eq.~(\ref{3})).
However, the $\mu$-dependence can be numerically large if
$i)$ the perturbative functions $R_1(y), R_2(y),$ ..\ are large, and/or
$ii)$ $\mu$ is very different from the typical energy scale $Q$
of the process (in this case the $\ln Q/\mu$-terms in Eq.~(\ref{3}) are
large and may spoil the convergence of the perturbative expansion).
For these reasons, assuming a well-behaved perturbative
expansion, $\mu$ should be set equal to $Q$ and $\mu$ variations (typically by
a factor of four) around this value can be used for estimating the theoretical
uncertainty due to uncalculated higher-order contributions.

Before discussing in detail jet measurements at LEP/SLC, I have to add a last
general comment about non-perturbative effects. Any hadronic obsvervable
has necessarely a non-trivial non-perturbative component because of the
hadronization of quarks and gluons. Therefore, in general we should write
\begin{equation}
\label{4}
R = R_{{\rm pert}}(\as(\mu), Q/\mu; y) + {\cal O}\!\left( \left(
%\frac{1}{Q} \right)^p \right) \;\;,
1/Q \right)^p \right) \;\;,
\end{equation}
where the last term represents the non-perturbative contribution. One of the
main properties of the jet observables, ensuing from their infrared and
collinear safety, is that the non-perturbative component is suppressed by some
inverse power of the energy as the energy increases (i.e. $p \geq 1$ in
Eq.~(\ref{4})). Therefore, in principle, the hadronization effects are
negligible at asymptotic energies. In practice, since we are dealing with
experiments at finite energy, we should try to estimate the size of the
non-perturbative corrections.

At present there are essentially two methods for estimating non-perturbative
contributions. The first method is based on the Operator Product Expansion
(OPE) [\ref{OPE}] and consists in relating the non-perturbative term
in Eq.~(\ref{4}) to the matrix elements of some local operators. These
matrix elements are not computable in perturbation theory but are universal,
in the sense that their value does not depend on the particular jet quantity
considered. Therefore one can estimate them from low-energy measurements
(or from lattice calculations) and evaluate the corresponding correction
to the jet observables.

The second method is based on Monte Carlo event generators
[\ref{MC}-\ref{MC4}].
The Monte Carlo generators used at LEP/SLC consist of two different
components. In the perturbative component, starting from the hard energy
scale of the process, one generates a parton shower describing quark and gluon
production according to (approximate) QCD matrix elements.
In the non-perturbative component, at a certain scale of the order of 1~GeV,
partons from the shower are converted into hadrons according to some
phenomenological hadronization model. The hadronization parameters are tuned
in order to reproduce the experimental data. Therefore, comparing jet
quantities as obtained by Monte Carlo simulations both from partons at the
end of the QCD shower and from particles after hadronization, one is able to
estimate the size of the non-perturbative corrections.

\pagebreak

\vskip 4.0mm
{\bf \noindent 3. Fully inclusive quantities}
\vskip 2.0mm

Fully inclusive quantities are jet observables depending on a single
momentum scale. Their perturbative expansion in Eq.~(\ref{3}) is thus a simple
power series in $\as$ with constant ($y$-independent) coefficients $R_n$.
Moreover, due to the highly inclusive nature of these quantities, they are
expected to be marginally affected by non-perturbative corrections
($p \geq 2$ in Eq.~(\ref{4})). For these reasons, these jet observables
are particularly suitable for $\as$ determinations. Two of them have been
investigated at LEP so far: the hadronic width of the $Z^0$ boson and the
hadronic branching ratio of the $\tau$ lepton.

\vskip 4.0mm
{\it \noindent 3.1 Hadronic branching ratio of the} $Z^0$ {\it boson}
\vskip 2.0mm

The hadronic branching fraction $R_Z$ of the $Z^0$ boson is defined by
the ratio of the hadronic and leptonic widhts:
\begin{equation}
\label{5}
R_Z=\frac{\Gamma_{\rm had}(M_Z)}{\Gamma_{\rm lep}(M_Z)} \;\;,
\end{equation}
and is computable in terms of electroweak and QCD corrections as follows
\begin{equation}
\label{6}
\Gamma_{\rm lep}(M_Z) = \Gamma_{\rm ew}^V + \Gamma_{\rm ew}^A \;\;,
\end{equation}
\begin{equation}
\label{7}
\Gamma_{\rm had}(M_Z)= \Gamma_{\rm ew}^V \,R_{QCD}^V + \Gamma_{\rm ew}^A
\, R_{QCD}^A \;\;.
\end{equation}
Here $\Gamma_{\rm ew}^{V,A}$ include all the known electroweak corrections
[\ref{Pas}] within the standard model.

The QCD contributions
$R_{QCD}^{V,A}$ are evaluated starting from the imaginary part of
$\Pi^{(i)}(Q^2)$, the correlation functions of the vector ($i=V$)
and axial ($i=A$) currents. These quantities are completely
known\footnote{The ${\cal O}(\as^3)$-singlet part of the axial current, which
was previously missing, has been computed recently [\ref{singlet}].}
up to next-to-next-to-leading order (NNLO) in perturbation theory
(i.e. to relative accuracy ${\cal O}(\as^3)$
with respect to the lowest-order approximation) [\ref{Gor},\ref{Kuhn}],
including all the relevant corrections due to finite quark masses
[\ref{mass1},\ref{mass2}].

In spite of these very accurate theoretical results, last year at the
EPS Conference
\linebreak
\noindent I pointed out [\ref{EPS}] that there were some discrepancies
in the actual numerical implementations of these calculations. The effect of
these discrepancies turned out to be of the same size as the expected
theoretical uncertainties.

This problem has been considered in a recent analysis by the authors
of Ref.~[\ref{Pas}]. They have shown that the numerical discrepancies are
eliminated after full updating of all the programs [\ref{BHM}]. Moreover,
they have provided  a simple effective (factorized) formula for $R_Z$ and
have carefully estimated its theoretical accuracy.

The QCD corrections $R_{QCD}^{V,A}$ to the vector and axial part of the
electroweak current are different. Therefore, in general, starting from
Eqs.~(\ref{6}) and (\ref{7}) one cannot write down a simple power series
in $\as$ for the ratio $R_Z$ in Eq.~(\ref{5}). Nonetheless, in the
proper range of electroweak parameters, one can derive an approximate
factorized expression for $R_Z$. The effective formula obtained in
Ref.~[\ref{Pas}] is\footnote{This formula is valid for massless leptons.
Corrections due to finite lepton masses can easily be taken into account
[\ref{Pas}].}
\begin{eqnarray}
\label{8}
&&R_Z \simeq R_0
\left[1+1.060 \frac{\as}{\pi}
%\right. \nonumber \\ &&\left.
+ 0.90
\left(\frac{\as}{\pi}\right)^2-15 \left( \frac{\as}{\pi}\right)^3 \right]
\;, \nonumber \\
&&R_0= 19.943 \;\;,
\end{eqnarray}
where $\as$ stands for $\as(M_Z)$. In the range $0.10 < \as < 0.15$,
Eq.~(\ref{8}) agrees with the full perturbative formula for $R_Z$ to an
accuracy better than $\Delta \as < 0.0001$.

Using the expression (\ref{8}), one can traslate theoretical uncertainties
on the value of $R_Z$ into corresponding uncertainties on $\as(M_Z)$.
The estimate in Ref.~[\ref{Pas}] gives
\begin{equation}
\label{9}
\Delta \as(M_Z)= \pm 0.002 \,({\rm ew}) \pm 0.002
\,({\rm QCD}) \; ^{+0.004}_{-0.003} \,(M_t, M_H) \;\;.
\end{equation}
The electroweak error (ew) is mainly due to electroweak corrections to
the $Z^0 \rightarrow b{\bar b}$ vertex. The QCD error is dominated by
unknown higher orders in perturbation theory and is estimated by varying
the renormalization scale $\mu$ within the range $M_Z/4 < \mu < M_Z$. Note
that non-perturbative corrections to $R_Z$ are negligible because of the
large value $M_Z \simeq 91$~GeV of the $Z^0$ mass: from the OPE analysis
the non-perturbative contributions turn out to be proportional to $(1/M_Z)^4$
(even including some additional corrections of the type $\Lambda^2/M_Z^2$, one
obtains $\Delta \as < 0.001$).

The dominant source of uncertainty in Eq.~(\ref{9}) is due to the values of
the masses of the top quark $(M_t)$ and Higgs boson $(M_H)$. The coefficients
in Eq.~(\ref{8}) refer to $M_t=150$~GeV and $M_H=300$~GeV, whilst the
corresponding errors in Eq.~(\ref{9}) are obtained by considering
variations in the range
$200\,{\rm GeV} > M_t > 100\,{\rm GeV} ,\,
1\,{\rm TeV} > M_H > 60\,{\rm GeV}$.

Including the data collected in the 1993 run, the updated LEP average for $R_Z$
is $R_Z=20.795 \pm 0.040$ [\ref{Schaile}]. From this value
and using Eq.~(\ref{8}) one gets:
\begin{equation}
\label{10}
\as(M_Z)=0.124 \pm 0.006 ({\rm exp.}) ^{+0.005}_{-0.004} ({\rm th.}) \;\;,
\end{equation}
where the experimental error is dominated by the event statistics.

This result is perfectly consistent with the value
$\as(M_Z)=0.126 \pm 0.005 ({\rm exp.}) \pm 0.002 (M_H)$ (and
$M_t ({\rm GeV}) = 173 \pm 13 ({\rm exp.}) \pm 19 (M_H)$)
obtained from a standard-model fit to the $Z^0$ lineshape and asymmetries at
LEP
[\ref{Schaile}]. However the latter is more dependent on the parameters of the
electroweak theory (in particular, $M_t$).

\vskip 4.0mm
{\it \noindent 3.2 Hadronic branching ratio of the} $\tau$ {\it lepton}
\vskip 2.0mm

An independent determination of $\as$ in NNLO can be obtained from the hadronic
width of the $\tau$ lepton. This quantity is indeed theoretically related
to the current correlation functions $\Pi^{(i)}(Q^2)$ via a simple momentum
sum rule [\ref{BNP}]. The
corresponding hadronic branching ratio is given by
\begin{equation}
\label{11}
R_{\tau}=\frac{B(\tau \rightarrow \nu_{\tau}+{\rm had.})}
{B(\tau \rightarrow \nu_{\tau} l {\bar \nu}_{l})}=R^{(0)} [1+
\delta_{\rm pert.} + \delta_{\rm non-pert.}] \;\;,
\end{equation}
where $R^{(0)}$ is the parton model value and the perturbative QCD correction
(as evaluated from the three-loop calculation of $\Pi^{(i)}(Q^2)$, described
in the previous subsection) is
\begin{equation}
\label{12}
\delta_{\rm pert.}= \frac{\as(M_{\tau})}{\pi} +
5.2 \left(\frac{\as(M_{\tau})}{\pi}\right)^2
+26.4 \left( \frac{\as(M_{\tau})}{\pi}\right)^3 \;.
\end{equation}

Unlike the case of $R_Z$, the non-perturbative contributions are potentially
sizeable at a scale as low as the $\tau$ mass $M_{\tau}=1.78$~GeV. Using
the OPE and estimating the relevant matrix elements by means of QCD sum rules,
the non-perturbative term in Eq.~(\ref{11}) is found to be [\ref{BNP}]
\begin{equation}
\label{13}
\delta_{\rm non-pert.}= -0.02 \pm 0.01 \;\;,
\end{equation}
which is quite a small contribution (although of the same order as the
${\cal O}(\as^3)$ term in Eq.~(\ref{12}), since $\as(M_{\tau}) \sim 0.3$).

Although this estimate is not completely unambiguous, at present we are
more confident on the value of $\delta_{\rm non-pert.}$. As a matter of fact,
the estimated value in Eq.~(\ref{13}) has been found consistent with the
hadronic mass spectrum of the $\tau$ lepton as measured by the ALEPH
Collaboration [\ref{ALEPH}].

The average value\footnote{This average value does not include two new
measurements of $R_{\tau}$ by the ALEPH and CLEO Collaborations [\ref{Webber}].
These results are {\em still} preliminary and differ each other by almost
three standard deviations.} of the results published by the four LEP
experiments [\ref{fer}] is $R_{\tau} = 3.617 \pm 0.034$. Using this value
and the theoretical predictions in Eqs.~(\ref{11})-(\ref{13}) one obtains
\begin{equation}
\label{14}
\as(M_{\tau})=0.36 \pm 0.02 ({\rm exp.}) \pm 0.04 ({\rm th.}) \;\;.
\end{equation}

The way of estimating the theoretical uncertainties on the value of
$\as$ from $\tau$ decay is still a matter of discussion
[\ref{Alt},\ref{Narison}]. The main points regard the validity of the OPE
close to the resonance region [\ref{Truong}] and possible corrections to the
(NNLO) perturbative running of $\as$ for scales as low as $M_{\tau}$
[\ref{Alt2}]. The theoretical error quoted in Eq.~(\ref{14}) takes into
account the uncertainty on $\delta_{\rm non-pert.}$ in Eq.~(\ref{13}), the
variation of the renormalization scale between 1~GeV and 3~GeV and the effect
of adding to Eq.~(\ref{12}) a contribution of order $\pm 100 (\as/\pi)^4$.

As expected from QCD, the value of $\as(M_{\tau})$ in Eq.~(\ref{14}) is
significantly larger than the value of $\as(M_Z)$  obtained from $R_Z$.
Extrapolating the result (\ref{14}) from the $\tau$ mass to the $Z^0$ mass
using the perturbative QCD running (see Eq.~(\ref{2})), one finds
\begin{equation}
\label{15}
\as(M_Z)=0.122 \pm 0.002 ({\rm exp.}) \pm 0.004 ({\rm th.}) \;\;,
\end{equation}
where the relative size of the errors is decreased because of the logarithmic
dependence of $\as$ on the energy.

\vskip 4.0mm
{\bf \noindent 4. Event shape distributions}
\vskip 2.0mm

The most detailed QCD tests performed so far at $\ee$ colliders are based on
studies of shape variables and jet cross sections.

The shape variables are global jet observables characterizing the structure
of the hadronic final
states. One of these variables is the thrust $T$, which is defined as
\begin{equation}
\label{16}
T= \nmax \frac{\sum_i |{\bf p}_i \cdot {\bf n}|}{\sum_i |{\bf p}_i|} \;\;,
\end{equation}
and thus maximizes the total longitudinal momentum (along the unit vector
$\bf n$) of the final-state particles $p_i$ in a given event. For a two-jet
event we have $T=1$, whilst a spherical event has $T=1/2$. Many other shape
variables can be defined [\ref{KN}], with the only constraint of being infrared
and collinear safe observables.

In the following I denote by $y$ a generic shape variable whose two-jet limit
corresponds to the region $y \rightarrow 0$ (for instance, $y=1-T$). The
corresponding shape-variable event fraction is defined by
\begin{equation}
\label{17}
R(\as(\mu), Q/\mu;y) = \int_0^y dy^{\prime} \frac{1}{\sigma}
\frac{d\sigma}{dy^{\prime}} \;\;,
\end{equation}
where $Q$ is the $\ee$ centre-of-mass energy and $\mu$ is
the renormalization scale.

The event fraction in Eq.~(\ref{17}) is computable in QCD perturbation theory
in the form of Eq.~(\ref{3}). Note that now the coefficients $R_1(y), R_2(y),
\dots$ are not $c$-numbers but functions of the actual value $y$ of the
shape variable. Therefore, unlike the case of the fully inclusive quantities,
by studying shape variable distributions one can not only measure $\as$
but also perfom more effective tests of the QCD matrix elements. The price
which
has to be paid is that, at comparable energy, the non-perturbative effects are
stronger than for the completely inclusive observables. Any time that the
degree
of inclusiveness is degraded, the effective power $p$, which controls the
energy
behaviour of the non-perturbative component (see Eq.~(\ref{4})), decreases.
According to the \naive\ expectation, the hadronization corrections to the
event
shapes are nominally of the order $\Lambda/Q$. This expectation is in agreement
with the non-perturbative effects estimated by Monte Carlo generators
and has been confirmed by calculations of renormalon contributions
[\ref{Webber2}].

\vskip 4.0mm
{\it \noindent 4.1 QCD studies to complete} ${\cal O}(\as^2)$
\vskip 2.0mm

The conventional approach for studying shape variables is based on the
comparison between data and NLO QCD calculations. For all the relevant shape
variables the ${\cal O}(\as)$ and ${\cal O}(\as^2)$ coefficients
$R_1(y)$ and $R_2(y)$ in Eq.~(\ref{3}) have been evaluated
numerically\footnote{In the case of the energy-energy correlation function,
the numerical disagreement among different calculations [\ref{KN},\ref{EEC}]
has been solved recently [\ref{Glov}] in favour of Ref.~[\ref{KN}].} by
Kunszt and Nason [\ref{KN}], using the two-loop matrix elements computed
in Ref.~[\ref{ERT}]. In these analyses the renormalization scale $\mu$
is usually set equal to $Q$. Note that, due to the explicit $y$-dependence
of the coefficients $R_n(y)$, by varying $\mu$ one can modify the shape in
$y$ of the QCD predictions. Therefore, a more
empirical approach is often considered: $\mu$ is left as a free parameter and
fitted to the data together with $\as$.

A summary [\ref{Mont}] of $\as$ determinations from NLO
calculations at LEP and SLC [\ref{fo}-\ref{SLD}]
is presented\footnote{The most recent (and complete) results from the SLD
Collaboration are not reported in Fig.~1. However, their inclusion does not
change the average value in Eq.~(\ref{18}).} in Fig.~1. All these measurements
are pretty consistent and the average value (taking into account experimental
and theoretical correlations) for $\as$ is
\begin{equation}
\label{18}
\as(M_Z)=0.119 \pm 0.001 ({\rm exp.}) \pm 0.006 ({\rm th.}) \;\;.
\end{equation}
The error is dominated by theoretical uncertainties (common to all the
measurements) due to hadronization
corrections and higher-order contributions. The hadronization
effects are estimated via Monte Carlo event generators by comparing the
corresponding results at parton and hadron level. The overall size of these
corrections is typically between 5 and $15\%$ (see, for instance Fig.~2).
The relative effect among
different Monte Carlo generators is instead of the order of few percent and is
usually assigned as the corresponding hadronization uncertainty.

Higher-order contributions are estimated from renormalization scale variations.
As a matter of fact, the peculiar feature of these QCD tests and $\as$
determinations to ${\cal O}(\as^2)$ is the following [\ref{fo},\ref{SLD}].
Far away from the
two-jet region, the NLO expression (\ref{3}) with a renormalization scale $\mu
\simeq Q$ gives good fits to the data. On the contrary, reasonable fits in the
two-jet region can be achieved only by using renormalization scales $\mu$ much
smaller than $Q$ (as small as few GeV!) and values of $\as$ smaller than
those obtained in the multi-jet region (Fig.~3).

\vspace*{13.5 cm}

\noindent {\bf Fig. 1}: Compilation of measurements of $\as(M_Z)$ from event
shapes, jet rates, energy correlations and scaling violations,
in ${\cal O}(\as^2)$, at LEP and SLC [\ref{Mont}].
\vskip 2.0mm

\pagebreak

\vspace*{10 cm}

\noindent {\bf Fig. 2}: DELPHI data on the heavy jet mass $M_h$
(the mass of the
heavier hemisphere with respect to the thrust axis) compared with the
${\cal O}(\as^2)$ QCD predictions (with a renormalization scale factor
$f \equiv \mu^2/Q^2 =0.25$). The hadronization corrections applied to the data
are shown below the distribution.
\vskip 2.0mm

\vspace*{7.8 cm}

\noindent {\bf Fig. 3}: Dependence of $\as(M_Z)$ (solid curves) and
$\chi^2$/d.o.f. (dashed curves) on $x_{\mu} \equiv \mu/Q$ for ${\cal O}(\as^2)$
fits to the OPAL data on thrust $(T)$, heavy jet mass $(M_H)$, total $(B_T)$
and
wide $(B_W)$ jet broadening.
\vskip 2.0mm

%\pagebreak

The reason for this strong scale dependence of the shape-variable event
fractions is that the corresponding perturbative functions
$R_n(y)$ in Eq.~(\ref{3}) are large in the two-jet region ($y \rightarrow
0$). For instance, in the case of thrust the actual calculation gives
[\ref{Bin}]
\begin{equation}
\label{19}
R_T(y=1-T) \t1 1-C_F\frac{\as}{\pi} \ln^2 (1-T) +
\frac{1}{2} \left( C_F\frac{\as}{\pi} \right)^2 \ln^4 (1-T)
+ {\cal O}(\as^3 \ln^6 (1-T))
\;.
%\;\;\; (T \rightarrow 1) \;.
\end{equation}
The double logarithmic contributions $\as \ln^2 (1-T), \,\as^2 \ln^4 (1-T)
\, \dots $ are due to the
brems\-strahlung spectrum of soft and collinear gluons. Although infrared and
collinear singularities cancel in jet observables upon adding real and
virtual contributions, in the two-jet limit real emission is strongly
inhibited. The ensuing mismatch of real and virtual corrections generates
logarithmically-enhanced terms which spoil the convergence of the perturbative
expansion in $\as$.

Observables dominated by two-jet configurations are thus affected by a large
and {\it systematic} theoretical uncertainty due to higher-order logarithmic
corrections. Reliable predictions can be obtained only computing these
corrections and, if possible, resumming them to all orders in $\as$.

\vskip 4.0mm
{\it \noindent 4.2 QCD studies using resummed calculations}
\vskip 2.0mm

A detailed understanding of logarithmically enhanced terms now
exists for many shape variables [\ref{thrust}-\ref{CDFW}], namely those for
which the small-$y$ logarithms $L=\ln 1/y$ {\it exponentiate} [\ref{CTTW}].
The shape variable event fraction $R(\as,y)$ can thus be written as follows
(in order to simplify the formulae I set $\mu=Q$)
\begin{equation}
\label{RCSig}
R(\as;y) = C(\as)\Sigma(\as,L) + D(\as;y)\; ,
\end{equation}
where
\begin{eqnarray}
\label{lnr}
C(\as) &=& 1 + \sum_{n=1}^\infty C_n \as^n \nonumber \\
\ln\Sigma (\as,y) &=& \sum_{n=1}^\infty\sum_{m=1}^{n+1} G_{nm} \as^n L^m
 \nonumber \\
&=& L \,g_1(\as L) + g_2(\as L) + \as \,g_3(\as L) + \cdots \;,
\end{eqnarray}
and $D(\as;y)$ vanishes as $y\to 0$ order by order in perturbation theory.
The word exponentiation refers
to the fact that the terms $\as^n L^m$ with $m > n+1$ are absent from
$\ln R(\as;y)$, whereas they do appear in $R(\as;y)$ itself. In the expression
(\ref{RCSig}) the singular $\ln y$ dependence is entirely included in the
effective form factor $\Sigma$.
The function $g_1$
resums all the {\em leading} contributions $\as^n L^{n+1}$, while $g_2$
contains the {\em next-to-leading} logarithmic terms $\as^n L^n$,
and $g_3$ etc.\ give the remaining {\em subdominant} logarithmic
corrections $\as^n L^m$ with $0<m<n$.

Equation (\ref{lnr}) represents an improved
perturbative expansion in the two-jet region. Once the functions $g_i$ have
been computed, one has a systematic perturbative
treatment of the shape distribution throughout the region of $y$ in which
$\as L\ltap 1$, which is much larger than the domain $\as L^2 \ll 1$ in which
the $\as$ perturbative expansion (\ref{3}) is applicable.
Furthermore, the resummed expression (\ref{lnr}) can be consistently matched
with fixed-order calculations. In particular, one can consider the
next-to-leading logarithmic approximation (NLLA) as given by the
functions $g_1$ and $g_2$ and
combine them with the ${\cal O}(\as^2)$ results in eq.~(\ref{3})
(after subtracting the resummed logarithmic terms in order to avoid double
counting), to obtain a prediction
(NLLA+${\cal O}(\as^2)$) which is everywhere at least as good as the
fixed-order result, and much better as $y$ becomes small.

\vspace*{20 cm}
\nopagebreak{
\noindent {\bf Fig. 4}: (a) Measured distributions of thrust $(\tau=1-T)$,
heavy jet mass $(\rho=M_H^2/Q^2)$, total $(B_T)$ and wide $(B_W)$
jet broadening,
compared with fits to the ${\cal O}(\as^2)$ and to the
resummed NLLA+${\cal O}(\as^2)$ calculations, with a renormalization scale
factor $\mu/Q =1$ in both cases.
%Sizes of
The (b) hadronization
%correction
and (c) detector corrections
%factors
applied to the data are also shown.}
\vskip 2.0mm

Extensive experimental studies based on NLLA+${\cal O}(\as^2)$ calculations
have been carried out during the last two years
[\ref{SLD},\ref{LEPres},\ref{Confres}].
As expected [\ref{thrust}] from the improved theoretical
accuracy of these predictions, it has been shown
that the
resummed calculations have a reduced dependence on the renormalization scale
and, in particular, remove the need to choose `unphysical' (very small)
renormalization scales in the two-jet region (see, for instance, Fig.~4).

\vskip 2.0mm\vspace*{11.5 cm}

\noindent {\bf Fig. 5}: Compilation of measurements of $\as(M_Z)$ from LEP and
SLC, using resummed NLLA+${\cal O}(\as^2)$ calculations [\ref{Mont}].
\vskip 4.0mm

A partial summary of $\as(M_Z)$ from resummed calculations is given in Fig.~5
[\ref{Mont}]. This summary does not include a very recent and detailed
analysis performed by the SLD Collaboration [\ref{SLD}]. Combining the new
SLD result ($\as(M_Z)=0.118 \pm 0.006$) with those of the LEP experiments, I
obtain the (correlated) average value
\begin{equation}
\label{22}
\as(M_Z)=0.122 \pm 0.002 ({\rm exp.}) \pm 0.005 ({\rm th.}) \;\;.
\end{equation}

This result is in good agreement with that obtained from analyses in
${\cal O}(\as^2)$
alone and has a comparable uncertainty. Note however that the theoretical
errors are treated differently. All the central values in Fig.~5 refer to the
same renormalization scale value $\mu=Q$ and scale uncertainties are evaluated
by varying $\mu$ by (approximately) a factor of four around $Q$. This range
includes the best-fit values for $\mu$: renormalization scales very different
from $Q$ are not only theoretically disfavoured but they also fail in
describing the data. Non-perturbative effects are again estimated from Monte
Carlo event generators. However, since modern Monte Carlo simulations operate
by generating parton configurations of arbitrarily large multiplicity
(typically
much larger than the maximum of four partons involved at order
${\cal O}(\as^2)$), they are better suited to estimating the hadronization
corrections for resummed calculations. In summary, using resummed
calculations there is much less freedom to define the central value of $\as$
and its theoretical uncertainty and a more consistent QCD picture emerges.

\vskip 4.0mm
{\it \noindent 4.3 Summary of measurements of} $\as$ {\it at} LEP {\it and} SLC
\vskip 2.0mm

The measurements of $\as$ at LEP and SLC can be summarized as follows (Fig.~6).
We have two largely independent determinations of $\as$ at the scale $M_Z$ from
$R_Z$ and event shape distributions in NLLA+${\cal O}(\as^2)$. They
respectively
give (see Eqs.~(\ref{10}) and (\ref{22})):
\begin{equation}
\label{A1}
\as(M_Z)=0.124 \pm 0.008 \;\;,
\end{equation}
\begin{equation}
\label{A2}
\as(M_Z)=0.122 \pm 0.006 \;\;.
\end{equation}
A third determination of $\as$ at the scale $M_{\tau}$ is provided by
$R_{\tau}$:
\begin{equation}
\label{A3}
\as(M_{\tau})=0.36 \pm 0.05 \;\;.
\end{equation}

\vspace*{8.5 cm}

\noindent {\bf Fig. 6}: Summary of measurements of $\as$ from $R_Z$ and
$R_{\tau}$ at LEP, and from event shape distributions at LEP and SLC.
\vskip 4.0mm

The two values of $\as(M_Z)$ in Eqs.~(\ref{A1}) and (\ref{A2}) are very
consistent with each other, thus proving that perturbative QCD is extremely
successful in describing hadronic interactions at the energy ${\sqrt s} \simeq
91$~GeV of the $Z^0$ boson mass. Moreover, as shown in Fig.~6, combining these
two determinations of $\as(M_Z)$  with that of $\as(M_{\tau})$ from $\tau$
decay, we have a strong evidence for the QCD running of $\as$ from LEP (and
SLC) data alone! Assuming the QCD running of Eq.~(\ref{2}), the value of
$\as(M_{\tau})$ in Eq.~(\ref{A3}) corresponds to $\as(M_Z)=0.122 \pm 0.005$.
Therefore, considering that the corresponding energy scales differ by almost
two orders of magnitude $(M_Z \simeq 50 \,M_{\tau})$, the agreement among these
determinations of $\as$ is a very significant test of QCD.

Let me also recall that the measurements of $\as$ at LEP/SLC are in good
agreement with those obtained from other processes. The world summary of $\as$
presented by S.~Bethke at the Montpellier Conference [\ref{Mont}] is reported
in
Fig.~7. The data points denoted by circles at the extremes of the plot
correspond to determinations performed at LEP/SLC.  Further evidence for the
QCD running is due to the measurements at intermediate values of the energy
$Q$,
which are obtained mainly from deep inelastic lepton-hadron scattering (DIS).
The values of $\as(M_Z)$ from DIS are slightly smaller than those from LEP/SLC
and lead to the consistent (and dominated by theoretical uncertainties) world
average $\as(M_Z)=0.117 \pm 0.006$ [\ref{Webber},\ref{BC}-\ref{Mont}].

\vspace*{10.2 cm}

\noindent {\bf Fig. 7}: Summary of measurements of $\as$ from different
processes and comparison with perturbative QCD expectations [\ref{Mont}].
%\vskip 4.0mm

\vskip 4.0mm
{\bf \noindent 5. Jet cross sections}
\vskip 2.0mm

A jet is qualitatively defined as a collimated spray of high-energy hadrons. It
is likely to be produced by hard scattering of partons and thus it can be
regarded as a universal signal of parton dynamics at short distances. However,
for the purpose of performing accurate quantitative studies,
one needs a precise
definition of jet. Essentially, one has to specify how low-energy particles
are assigned to jets, in order to have infrared finite cross sections.

\vskip 4.0mm
{\it \noindent 5.1 Jet algorithms}
\vskip 2.0mm

The standard jet definition in $\ee$ annihilation [\ref{JADE},\ref{KN}] amounts
to introducing a dimensionless resolution variable $y_{ij}=d_{ij}/Q^2$ for
every pair $(i,j)$ of particles (jets). Then the particles (jets) with the
minimum $y_{ij}$ are merged until a fixed resolution $\ycut$ is reached
($y_{ij} > \ycut$). The final-state particles in each event are therefore
classified in a well defined number of jets, depending on the resolution
$\ycut$. The main feature of this definition is that the corresponding
iterative procedure of clustering type provides an unambiguous and exhaustive
assignment of particles to jets.

Several jet clustering algorithms for $\ee$ annihilation are available
[\ref{KN},\ref{BKSS}]. Different jet algorithms are specified by the
definition of the dimensionful resolution variable $d_{ij}$. As a result of
many thoretical and phenomenological investigations carried out in the last
few years [\ref{CDOTW},\ref{BKSS}-\ref{DO}], the theoretically favoured
resolution variable turns out to be:
\begin{equation}
\label{dkt}
d_{ij}^{(k_{\perp})}= {\rm min} \;2(E_i^2,E_j^2) (1-\cos \theta_{ij}) \;\;.
\end{equation}
Here $E_i$ and $E_j$ are the particle (jet) energies in the $\ee$
centre-of-mass
frame and $\theta_{ij}$ is their relative angle. This resolution variable
reduces to the minimal relative transverse momentum
$k^2_{\perp ij}$ in the
limiting case of small relative angles ($\theta_{ij} \rightarrow 0$). For this
reason the algorithm is known as $k_{\perp}$-algorithm\footnote{This algorithm
was first discussed at the Durham Workshop on Jet studies at LEP and HERA,
December 1990 [\ref{DURHAM}], and is sometimes referred to as the Durham
algorithm.}.

In order to motivate the preference for the $k_{\perp}$-algorithm,
let me compare\footnote{A more detailed discussion can be found in
Ref.~[\ref{jettop}].} its features with those of an older jet definition as
given by the JADE algorithm [\ref{JADE}]. In the latter the jet resolution
variable is essentially the invariant mass
$d_{ij}^{(J)}=2E_iE_j(1-\cos \theta_{ij}) \simeq (p_i+p_j)^2$ of the pair of
particles (jets).

The resolution variables $d_{ij}^{(k_{\perp})}$ and
$d_{ij}^{(J)}$ both vanish in the soft limit $E_i,E_j \rightarrow 0$ (the
two algorithms are thus infrared safe) but they behave differently for small
(and finite) particle energies. It follows that they treat soft radiation in
a different
way. In particular, since $d_{ij}^{(J)}$ depends on the product $E_iE_j$, the
JADE algorithm prefers to merge soft particles first, even if they are far
apart in angle. In the case of the $k_{\perp}$-algorithm, the resolution
variable $d_{ij}^{(k_{\perp})}$ is instead diagonal with respect to particle
energies (i.e. the product $E_iE_j$ is replaced by $E_i^2$ or $E_j^2$). Hence,
soft particles are merged with the energetic particle closest in angle. The
different jet classification is clearly shown by the L3 event reported in
Fig.~8 [\ref{Heb}]. One can say that the JADE algorithm induces strong
{\it attractive
kinematic} (due to the jet definition and independent of the underlying
dynamics) {\it correlations} among soft particles. On the contrary, the
$k_{\perp}$-algorithm, where these correlations are absent, leads to the
several advantages: $i)$ it avoids a non-intuitive classification of events
and an unnatural assignment of particles to jets (soft and wide-angle jets);
$ii)$ it reduces the size of the non-perturbative corrections due
to the hadronization process
(since soft particles are merged with the energetic
particle closest in angle, soft fragmentation products are likely to be
assigned to the `parent' jet); $iii)$ theoretical calculations in
perturbative QCD are more reliable, because multi-parton kinematics does not
dominate multi-parton dynamics in higher perturbative orders.

The last point is particularly evident in the small-$\ycut$ region, where
the QCD perturbative expansion of jet cross sections is dominated by large
double logarithmic corrections of the type $\as^n \ln^m \ycut \,(m \leq 2n)$.
These contributions, whose origin is due to the emission of soft and collinear
gluons (as in the case of event shapes in the two-jet region), can be resummed
to all orders in $\as$ if jets are defined using the $k_{\perp}$-algorithm. On
the contrary, in the case of the JADE algorithm, the attractive kinematic
correlations among soft particles prevent the implementation of the resummation
procedure [\ref{BS}].

\vspace*{5 cm}

\noindent {\bf Fig. 8}: A 3-jet event as `seen' by (left) the JADE algorithm
and (right) the $k_{\perp}$-algorithm.
\vskip 4.0mm

One more attractive feature of the $k_{\perp}$-algorithm is its applicability
to processes involving initial-state hadrons (i.e. deep inelastic lepton-hadron
scattering, photoproduction processes and hadron-hadron collisions). The
generalization to hadron collisions of $\ee$ clustering algorithms is by no
means trivial, because one has to face the problem of dealing with the soft
remnants of the incoming hadrons and factorize them from high-$p_{\perp}$ jets
produced by hard scattering of partons. As proposed in Ref.~[\ref{CDW}],
resolution variables of transverse-momentum type are particularly suitable for
this purpose and, actually, a $k_{\perp}$-clustering algorithm for hadron
collisions has been set up in Refs.~[\ref{CDW},\ref{kthad}]. Its use for
studying jet physics at HERA and at the Tevatron collider may offer some
advantages [\ref{CDW}-\ref{Sey}] with respect to standard cone algorithms
[\ref{SA}].

\vspace*{9 cm}

\noindent {\bf Fig. 9}: Comparison between ALEPH data on the differential
two-jet rate $D_2(\ycut=y_3)$ and resummed+${\cal O}(\as^2)$ calculations
corrected (solid curve) for hadronization effects.
%\vskip 4.0mm

\vskip 4.0mm
{\it \noindent 5.2 Jet rates and multiplicities}
\vskip 2.0mm

In the last few years, the
$k_{\perp}$-algorithm has replaced the original JADE algorithm in most
of the jet analyses carried out in $\ee$ annihilation,
including determinations of $\as$ and
studies of jet topology and coherence effects.

The basic jet measurements one can consider are the $n$-jet rates
$R_{n-jet}=\sigma_{n-jet}/\sigma_{TOT}$, defined by the ratio between the
cross section $\sigma_{n-jet}$ for producing $n$ jets (at the resolution scale
$\ycut$) and the total cross section $\sigma_{TOT}$. In Fig.~9, the ALEPH
data [\ref{LEPres}] on the differential two-jet rate
$D_2(\ycut)=dR_{2-jet}(\ycut)/d\ycut$ are compared with the corresponding
QCD predictions in ${\cal O}(\as^2)$ and including the all-order resummation
of  the logarithmic contributions $\as^n \ln^{2n}\ycut$ and
$\as^n \ln^{2n-1}\ycut$ [\ref{CDOTW}]. The value of $\as$ extracted from this
analysis contributes an entry in the list reported in Fig.~5.

As a further example of jet studies using the $k_{\perp}$-algorithm,
let me consider the average jet multiplicity
$\VEV{n_{jet}}=\sum_{n \geq 2} n \,R_{n-jet}$.
%\sigma_{n-jet}/\sigma_{TOT}$.
This measurement can be considered as an alternative
QCD test with respect to the mean multiplicity of hadrons produced in $\ee$
annihilation. In the case of the hadron multiplicity one is interested in
its dependence on the $\ee$ centre-of-mass energy (see, for instance, Sect.~6.2
in Ref.~[\ref{Heb}]). Here, at a fixed centre-of-mass energy, one can
instead study the jet multiplicity as a function of the jet resolution $\ycut$.

\vspace*{8 cm}

\noindent {\bf Fig. 10}: L3 data on the average jet multiplicity
$\VEV{n_{jet}}$
compared with NLLA+${\cal O}(\as^2)$ predictions.
\vskip 4.0mm

An appealing feature of $\VEV{n_{jet}}$ is that, unlike the
hadron multiplicity, it is an infrared safe observable and thus its absolute
normalization (and not only its dependence on the energy) is computable in
perturbation theory. Moreover, the increase of the jet multiplicity is highly
sensitive to non-abelian effects,
i.e. to multiple jet production by gluon cascades. This effect is particularly
evident in the small-$\ycut$ region where the QCD calculation, to leading
accuracy in the double logarithmic expansion parameter $a\equiv (\as/2\pi)
\ln^2
\ycut$, predicts:
\begin{equation}
\label{njet}
\VEV{n_{jet}} - 2 \smally C_F a \left[1+\frac{1}{12} C_A a +
\frac{1}{360} C_A^2 a^2 + \cdots \right] \;\;.
\end{equation}
Note that in this approximation the deviation of Eq.~(\ref{njet}) from the
double logarithmic behaviour $\as \ln^2 \ycut$ is entirely due to non-abelian
(i.e. proportional to powers of the gluon charge $C_A=N_c$) contributions. The
QCD predictions for $\VEV{n_{jet}}$ in NLLA+${\cal O}(\as^2)$ [\ref{CDFW}]
have been
succesfully tested by the L3 and OPAL Collaborations [\ref{LEPres}] (see
Fig.~10).

\vskip 2.0mm
{\it \noindent 5.3 Other jet studies}
\vskip 2.0mm

The gauge structure of QCD is completely determined by the fact that the
gauge group is $SU(N_c)$ with $N_c=3$ colours, and that the quarks belong
to the fundamental representation. This fixes completely the relative magnitude
of all the couplings of the theory. Roughly speaking, to the splitting
processes $q \rightarrow qg, \,g \rightarrow gg$ and $g \rightarrow q{\bar q}$,
one can associate splitting probabilities respectively proportional to
$C_F= (N_c^2-1)/2N_c, \, C_A=N_c$ and $T_R=1/2$. Events with more than four
final-state jets in $\ee$ annihilation give access to all the relevant
splitting
processes and thus multijet events offer the possibility of performing very
detailed tests of the gauge structure of the theory.

Angular correlations within four-jet events have been investigated by the LEP
Collaborations [\ref{colour}]. In these investigations, the values
of the colour
factors $C_F, \, C_A, \,T_R$ in the two-loop matrix elements [\ref{ERT}] are
left as free parameters and fitted to the LEP data.  The results of the fits
are
consistent with the colour gauge group $SU(3)$ whilst the only other groups
with three quark colours (namely, $SO(3)$ and the abelian toy model introduced
in Ref.~[\ref{Kra}]) are excluded. Unfortunately, since only ${\cal O}(\as^2)$
matrix elements are at present available for these four-jet studies, the
ensuing
calculations are just LO predictions. Therefore it is very difficult to assess
the theoretical accuracy of these analyses.

Many more studies of jet properties have been performed at LEP and SLC.
Among them, the results, which have been recently reported
[\ref{qgjet},\ref{ALEPHg}] on the
evidence for differences between quark and gluon jets, are
particularly relevant. This and other jet topics
are reviewed elsewhere [\ref{Heb},\ref{Gat},\ref{Webber}]. In the next section,
I discuss in some detail a study of the internal structure of jets based
on the measurement of the subjet multiplicity inside jets. I think this
measurement is quite interesting for addressing issues like QCD coherence and
properties of quark and gluon jets.

\setcounter{footnote}{0}
\renewcommand{\thefootnote}{\fnsymbol{footnote}}

\vskip 4.0mm
{\bf \noindent 6. Sub-jet multiplicity and QCD coherence}

\vskip 2.0mm
{\it \noindent 6.1 Quark and gluon jets}
%\vskip 2.0mm

One of the main features of perturbative QCD is the different colour charge
carried by quarks and gluons. The bremsstrahlung spectra for the emission of
a soft gluon from a hard quark and from a hard gluon differ only in their
relative normalization. The emission probability from a quark (or an antiquark)
is proportional to the square $|{\bf T}_q|^2=C_F=4/3$ of the (non-abelian)
quark charge\footnote{The quark and gluon colour charges ${\bf T}_q$ and
${\bf T}_g$ are $SU(3)$ colour matrices respectively in the fundamental and
adjoint representations.} ${\bf T}_q$, whilst the emission probability from a
gluon is proportional to the square $|{\bf T}_g|^2=C_A=3$ of the gluon
charge ${\bf T}_g$. Among the quantities which are easily measurable in the
experiments, the average particle (parton) multiplicity is the most sensitive
to
the radiation of soft gluons. If we compare the particle multiplicities
${\cal N}_g$ and ${\cal N}_q$ in a gluon and a quark jet, we may thus expect
a ratio ${\cal N}_g/{\cal N}_q$ which is approximately unity at low energies
and increases with the energy. At asymptotic energies (in practice, as soon as
${\cal N}_g, {\cal N}_q \gg 1$) this ratio should be equal to the ratio of the
(squares of the) colour charges, ${\cal N}_g/{\cal N}_q \simeq C_F/C_A=9/4$,
i.e. larger than a factor of two.

This \naive\ expectation has two main problems, besides possibly large
subasymptotic corrections [\ref{Mue}]. One problem is related to the
hadronization effects. The hadron multiplicity is not an infrared safe
observable and thus it cannot be computed in QCD perturbation theory. In other
words, we have poor theoretical control over the non-perturbative contribution
to the ratio of the hadron multiplicities: some systematic difference in
the hadronization of quark and gluon jets could significantly contribute to
a deviation from the asymptotic multiplicity ratio [\ref{ALEPHg}].

The other problem is related to the definition and identification of quark and
gluon jets. As discussed at the beginning of Sect.~5, the quantitative
definition of jet is a highly non-trivial issue, which can be solved only by
introducing jet-finding algorithms. The problem is amplified in the case of
a distinct definition of quark and gluon jets because, although we observe
jet-like objects, quarks and gluons are not experimentally accessible. Thus,
quark jets and gluon jets can be singled out by comparing different processes
in which only quarks or gluons contribute at the level of the \naive\ parton
model. Alternatively, in a given process one can anti-tag a gluon jet by
identifying heavy-quark decays [\ref{qgjet},\ref{ALEPHg}].
No matter how quark and gluon jets
are defined, their definition is theoretically biased by our underlying
partonic
picture. More quantitatively, we can say that quark and gluon jets cannot be
defined in a universal way to an accuracy better than ${\cal O}(\as)$,
whether due to the dependence on the process or the dependence
on the observable.
Moreover, although nominally of ${\cal O}(\as)$, this dependence can be
numerically large because typically enhanced by logarithmic coefficients (as
discussed in Sect.~5.1, even the double logarithmic contributions
$\as \ln^2 \ycut$ to the jet rates do depend on the jet definition).

Some years ago [\ref{sub}], we propose a method aimed at investigating
differences between quark and gluon jets by overcoming the problems discussed
so far. The idea is to select two-jet and three-jet event samples with the
$k_{\perp}$-algorithm at a fixed value of the jet resolution parameter
$\ycut \equiv y_1$. Then the $k_{\perp}$-algorithm is
again applied by clustering
subjets with a smaller resolution $\ycut \equiv y_0 \leq y_1$ and counting the
number of subjets in each jet sample.

Within this approach the hadronization corrections are under better control
because hadrons are replaced by subjets, which are infrared and collinear safe
objects. Moreover, $\ee$ annihilation is a point-like source of $q{\bar q}$
events which undergo fragmentation through the splitting process $q{\bar q}
\rightarrow q{\bar q}g$. Therefore, comparing the two-jet and three-jet
samples,
one has access respectively to the properties of a mixture of $q{\bar q}$ jets
and $q{\bar q}g$ jets without further specification and identification of
quark and gluon jets.

In particular, one can consider the ratio of the average subjet multiplicities
$M_3$ and $M_2$ in three-jet and two-jet events. By fixing $y_1$ and decreasing
$y_0$, the ratio $M_3/M_2$ is expected to vary in the range:
\begin{equation}
\label{32}
\frac{3}{2} \leq \frac{M_3}{M_2} \rightarrow \frac{{\cal N}_q +
{\cal N}_{\bar q} + {\cal N}_g}{{\cal N}_q + {\cal N}_{\bar q}} \simeq
\frac{2C_F+C_A}{2C_F} = \frac{17}{8} \;\;.
\end{equation}
The value $3/2$ on the left-hand side of Eq.~(\ref{32}) corresponds to
$y_0=y_1$
and is simply due to kinematics (at $y_0=y_1$  subjets and jets coincide and
there are exactly 3 jets in the three-jet sample and 2 jets in the two-jet
sample). The right-hand side of Eq.~(\ref{32}) is instead approached at $y_0
\ll y_1$. Actually, the value 17/8 is that \naively\ expected as $y_0/y_1
\rightarrow 0$ on the basis of the (asymptotic) charge counting rule
${\cal N}_g/{\cal N}_q = {\cal N}_g/{\cal N}_{\bar q} \simeq C_A/C_F$.

The theoretical [\ref{sub}] and experimental [\ref{23jet}] analyses show that
this \naive\ expectation is not only numerically violated but completely
misleading from a physical viewpoint. In any physical process, jets are not
produced independently because of colour conservation. This leads to {\em
coherence effects} that, in the case of the subjet multiplicities, are
responsible for a value of the ratio $M_3/M_2$  which remains well below 3/2
(which is the value at the kinematic boundary and not the asymptotic value!)
for most of the $y_0$ range [\ref{sub}].

The perturbative QCD predictions obtained in Ref.~[\ref{sub}] include the
resummation of the leading and next-to-leading logarithmic contributions
$\as^n L^{2n}$ and $\as^n L^{2n-1}$ ($L$ standing for both
$L_1= - \ln y_1$ and $L_0= - \ln y_0$) to all orders in $\as$.
The actual calculation is quite involved,
so that, in the following, I limit myself to presenting the final results and
to
providing their qualitative interpretation.

\vskip 4.0mm
{\it \noindent 6.2 Two-jet and three-jet events: $\Delta \eta$-rapidity
profiles}
\vskip 2.0mm

Considering jets produced at the $\ee$ centre-of-mass energy $Q$, let me
introduce the transverse-momentum scales $Q_1= {\sqrt {y_1}} Q$ and
$Q_0= {\sqrt {y_0}} Q$. Because of the definition (\ref{dkt}) for the
resolution
variable, jets selected by the $k_{\perp}$-algorithm at $\ycut = y_1$ have a
natural angular size $\theta_J \simeq 2Q_1/E_J$, $E_J$ being the jet energy
(Fig.~11a). This angular size identifies the jet core, i.e. the fragmentation
region of the jet: all the subjets (or particles), which are inside a cone of
aperture $\theta_J$ around the jet axis, belong to the jet. However, each jet
$J$ contains also all the subjets which are relatively soft (with energy much
smaller than $E_J$) and are produced at an angle $\theta$ (with respect to the
jet axis) in the region $\theta_J \ltap 2 \theta \ltap \theta_{J,J'}$,
$\theta_{J,J'}$ being the angular distance to the nearest jet $J'$ (Fig.~11).
This region can be called `interjet' region since its angular extension depends
on the topology of the multijet configuration.

Let me thus consider the (pseudo-)rapidity profile $dN/d\eta$ of a jet, that
is,
the number of subjets as a function of their pseudorapidity $\eta= - \ln \tan
(\theta/2)$ with respect to the jet axis.

A two-jet event with resolution $\ycut=y_1$ consists of the subjets radiated
by a leading quark and a leading antiquark (Fig.~11a). In the fragmentation
region of the quark jet the rapidity profile increases up to a saturation value
and then produces a rapidity plateau in the interjet region (Fig.~12). The
situation is perfectly symmetric in the antiquark hemisphere and the result
for the subjet multiplicity in the two-jet event is [\ref{sub}]:
\begin{equation}
\label{m2}
M_2(Q_0,Q_1;Q) = 2 \left\{ {\cal N}_q(Q_0,Q_1) + C_F \,H(Q_0,Q_1) \,\ln
\frac{Q}{Q_1} \right\} \;\;.
\end{equation}

The first term on the right-hand side of Eq.~(\ref{m2}) represents the
`intrinsic'
multiplicity of the quark and antiquark jets, coming from their natural
angular regions. This contribution is indeed obtained by integrating the
rapidity profile in Fig.~12 over the fragmentation regions.

The subjet multiplicity ${\cal N}_q(Q_0,Q_1)$ is equal to $\; \half
\VEV{n_{jet}(\ycut=Q_0^2/Q_1^2)}$,
$\left\langle n_{jet}(\ycut = \right.$
$\left. Q_0^2/Q^2) \right\rangle$
%$\VEV{n_{jet}(\ycut=Q_0^2/Q^2)}$
being
the total jet multiplicity discussed in Sect.~5.2. As a matter of fact,
if $y_1=1$ (i.e. $Q_1=Q \simeq E_J/2$) the fragmentation region extends up to
$\eta=0$ and all the $\ee$ events are identified as two-jet events.

The second term on the right-hand side of Eq.~(\ref{m2}) gives the extra
interjet multiplicity between the jet cores. Note that each jet contributes
to a rapidity plateau whose length and height are respectively
$\eta_1= \ln 2E_J/Q_1 \simeq \ln Q/Q_1$ and $C_F \,H(Q_0,Q_1) \simeq
\partial {\cal N}_q(Q_0,Q_1)/\partial \ln Q_1$. The height of the plateau is
proportional to the square $|{\bf T}_q|^2=C_F$ of the colour charge of the
leading quark.

\vspace*{7 cm}

\noindent {\bf Fig. 11}: (a) A two-jet event in $\ee$ annihilation.
The subjets $k_1$ and $k_2$ are produced respectively in the fragmentation
region (I) and in the interjet region (II). (b) Typical kinematic
configuration of a three-jet event: (A), (B), and (C) denote the angular
regions described in the text.
\vskip 4.0mm

\vspace*{5.5 cm}

\noindent {\bf Fig. 12}: Rapidity profile of a quark (antiquark) jet in a
two-jet event: (I) fragmentation and (II) interjet regions
%$(\eta_i=\ln 2E_J/Q_i \simeq \ln Q/Q_i, \, i=0,1)$.
$(\eta_0=\ln 2E_J/Q_0 \simeq \ln Q/Q_0,
\, \eta_1=\ln 2E_J/Q_1 \simeq \ln Q/Q_1)$.
%\vskip 4.0mm

Let me then consider three-jet events. When $y_1 \ll 1$ (as appropriate for
a calculation to logarithmic accuracy in $\ln y_1$) a three-jet event has the
typical kinematic configuration in Fig.~11b. An antiquark (quark) jet recoils
from a quark (antiquark) and a gluon jet which are produced at a small relative
angle $\theta_{qg}$ in the opposite hemisphere. The rapidity profile of the
antiquark jet is similar to that in the two-jet sample. The situation
is instead
different in the hemisphere with two jets. One has to examine subjet production
in three different angular regions (Fig.~11b): (A) at small angle
$\theta$ with respect
to the quark-jet axis $(\theta < \theta_{qg})$; (B) at small angle $\theta$
with respect to the gluon-jet axis $(\theta < \theta_{qg})$; (C) at large
angle $\theta$ with respect to both the quark-jet and the gluon-jet axis
$(\theta > \theta_{qg})$.

\vspace*{15.5 cm}

\noindent {\bf Fig. 13}: Rapidity profiles for the angular regions (A), (B)
and (C) (see text) of the quark+gluon jet hemisphere in three-jet events
$(\eta_{qg}= -\ln \tan \theta_{qg}/2)$.
\vskip 4.0mm

The angular regions (A) and (B) include the jet cores and lead to rapidity
profiles (Fig.~13) which are similar to that observed in two-jet events.
The only
difference between the quark jet (A) and the gluon jet (B) is that the latter
has a larger multiplicity ${\cal N}_g(Q_0,Q_1)$ coming from its fragmentation
region\footnote{The gluon multiplicity ${\cal N}_g(Q_0,Q)$ is equal to one half
of the total jet multiplicity one could measure in quarkonium decay into two
gluons.} and, correspondingly, a higher rapidity plateau. The height of the
rapidity plateau of the gluon jet is $C_A \,H(Q_0,Q_1) \simeq
\partial {\cal N}_g(Q_0,Q_1)/\partial \ln Q_1$, i.e. proportional to the
square $|{\bf T}_g|^2=C_A$ of the gluon charge.

Subjets produced in the interjet region (C) also contribute to a rapidity
plateau (Fig.~13).
However, since their emission angle $\theta$ is much larger than
$\theta_{qg}$, they cannot resolve two distinct jets. The height of the
plateau is thus  proportional to the square $|{\bf T}_q + {\bf T}_g|^2$ of the
total colour charge of the quark and gluon jets. Because of colour conservation
${\bf T}_q + {\bf T}_g = - {\bf T}_{\bar q}$ (${\bf T}_{\bar q}$ being the
colour charge of the antiquark jet recoiling in the opposite hemisphere), and
hence, the height of the plateau is proportional to $|{\bf T}_q + {\bf T}_g|^2
=|- {\bf T}_{\bar q}|^2 = C_F$ !

\vspace*{8 cm}

\noindent {\bf Fig. 14}: $\Delta \eta$-rapidity profiles for (a) two-jet
and (b) three-jet events $(L_i= -\ln y_i, \, i = 0,1)$. The suppression at
small $\Delta \eta$ in (b) is due to QCD coherence.
\vskip 4.0mm

The results of this discussion on three-jet events can be combined in the
rapidity profile of Fig.~14b. Here $\Delta \eta =\eta -\ln (2E_J/Q_0)$, where
$\eta$ is the subjet pseudorapidity with respect to the axis of the parent jet
and $E_J$ is the energy of the parent jet. Therefore the $\Delta \eta$-rapidity
profile is experimentally measurable without identifying quark and gluon jets.
Integrating the $\Delta \eta$-rapidity profile in Fig.~14b, one obtains
[\ref{sub}]:
\begin{equation}
\label{m3}
M_3(Q_0,Q_1;Q) = M_2(Q_0,Q_1;Q) + {\cal N}_g(Q_0,Q_1) + C_A \,H(Q_0,Q_1)
\, {\VEV{\eta}}_g \;\;.
\end{equation}

The term $M_2$ on the right-hand side of Eq.~(\ref{m3}) comes from the
contribution of the antiquark jet and from the contributions of the angular
regions (A) and (C) (which add themselves to reconstruct an entire quark-jet
rapidity profile as in Fig.~12) in the opposite
hemisphere. The remaining terms on the right-hand side of Eq.~(\ref{m3}) are
due to the additional gluon jet. Like the quark and antiquark jets, the gluon
jet provides an intrinsic and an interjet contribution. However, the latter is
restricted to the angular region $\theta_{qg} > \theta > Q_1/E_J$ around the
gluon jet axis. Thus the length of the pseudorapidity plateau is $\ln q/Q_1$,
where $q= {\rm min} (E_q \theta_{qg}, E_g \theta_{qg})$ is the transverse
momentum at which the gluon and quark jets are resolved. In Eq.~(\ref{m3}),
this
length is averaged over $q$ (${\VEV{\eta}}_g \equiv \VEV{ \ln q/Q_1}$) with the
appropriate three-jet cross section.

In Fig.~14a, I have also reported the $\Delta \eta$-rapidity profile of the
two-jet sample, in order to make easier the comparison with the
corresponding profile of the three-jet sample. We can see that the additional
gluon  jet contribution in Fig.~14b, although asymptotically higher by the
canonical factor $C_A/C_F=9/4$, is always shorter than those of the quark jets
if $y_1 \ll 1$. In this case the angle between the quark and the gluon jet is
very small and the gluon jet is strongly squeezed. This suppression of the
$\Delta \eta$-rapidity profile in the central region is due to QCD coherence,
that is, to destructive interference of the quark and gluon jets. In this
large-angle region, the quark and gluon jets act coherently with a
colour-charge factor of  $|{\bf T}_q + {\bf T}_g|^2 =|- {\bf T}_{\bar q}|^2
= C_F$, which is much smaller than $|{\bf T}_q|^2 + |{\bf T}_g|^2 = C_F + C_A$,
the colour-charge factor that one would consider in the case of independent
fragmentation of the two jets.

A general lesson one can learn from this discussion is that quark and gluon
jets do not live independently. Because of QCD coherence (colour conservation)
their properties depend on their production mechanism and on
the actual selection procedure.

\vskip 4.0mm
{\it \noindent 6.3 Multiplicities in two- and three-jet events}
\vskip 2.0mm

Figure 15 shows results on the ratio $M_3/M_2$ from LEP [\ref{23jet}]. As a
consequence of the squeezing of the gluon jet due to QCD coherence, this ratio
is predicted to fall off near the kinematic limit $y_0 \simeq y_1$ [\ref{sub}].
This prediction is in good agreement with the data.  Also shown is
the prediction of a toy model which does not take into account coherent
effects.
In this toy model, where jets contribute according to their energies and colour
charges without interference, the multiplicity ratio rises monotonically
towards
its asymptotic value, in disagreement with the experiments. As a further
check on
the origin of the decrease in the ratio $M_3/M_2$, one can measure its slope at
$y_0=y_1$ as a function of $y_1$. According to QCD coherence, by increasing
$y_1$ the angle $\theta_{qg}$ increases, thus reducing the region in which the
quark-gluon destructive interference takes place. The slope of $M_3/M_2$ is
thus predicted to decrease at large $y_1$, in agreement with the data
[\ref{23jet}].

\vspace*{6 cm}

\noindent {\bf Fig. 15}: L3 data on the ratio $M_3/M_2$ compared with the
QCD predictions in Eqs.~(\ref{m2}),(\ref{m3}) (solid curve) and with the
predictions of an incoherent toy model (dotted curve).
%\vskip 4.0mm

The measurement of the subjet multiplicity in two-jet and
three-jet events can be definitely considered a detailed test of (perturbative)
QCD coherence. Since it is based on infrared and collinear safe observables
which do not require any identification of the gluon jet, this test of
coherence is
certainly more `model independent' than the celebrated string effect
[\ref{string1}-\ref{string3}].

\vspace*{16 cm}

\noindent {\bf Fig. 16}: OPAL data on the subjet multiplicity of two-jet
events compared with different Monte Carlo models. The fractional difference
between models and data is also shown.
\vskip 4.0mm

The perturbative QCD predictions for $M_3/M_2$ are confirmed by the LEP data up
to $y_0 \simeq 5 \cdot 10^{-5}$. This value of $y_0$ corresponds to a minimal
transverse momentum $Q_0 \simeq 2$~GeV at LEP energies. It is thus evident that
perturbative QCD fails at smaller values of $y_0$ because the hadronization
effects dominate. On the contrary, Monte Carlo event generators can be used for
investigating the development of a jet by varying $y_0$ from the formation
scale
($y_0=y_1$) to the scale at which individual hadrons are resolved as subjets
($y_0 \sim 10^{-5}$ at $Q = M_Z$). The measurements of subjet properties
performed at LEP have been compared with different Monte Carlo models
[\ref{23jet}].
These include generators like JETSET [\ref{MC}], HERWIG [\ref{MC2}], ARIADNE
[\ref{MC3}], which implement coherence effects in the parton shower evolution,
and generators like COJETS [\ref{MC4}], which are based on independent
fragmentation. As shown by the representative study in Fig.~16, all the models
describe well the data. This is not surprising: the various (hadronization)
parameters have been tuned to describing many hadronic observables and such a
tuning can mask the dynamics differences among the models. However, it is
interesting to note that, although the agreement with the data is always very
good for $y_0 \ltap 10^{-5}$ (i.e. at scales where hadronization effects
dominate), the Monte Carlo generators with no QCD coherence perform worse
at the
true subjet level ($y_0 \gtap 10^{-5}$). Typically, the incoherent models
produce too much radiation and too many subjets in this phase space region.
This analysis is a further evidence in favour of coherence and shows that the
study of subjets can be a useful tool for investigating the transition from
the perturbative to the non-perturbative phase of QCD.

\vskip 4.0mm
{\bf \noindent 7. Outlook}
\vskip 2.0mm

As discussed in the previous sections, with the advent of the high-energy and
high-statistics experiments at LEP/SLC and of accurate theoretical
calculations,
`high-precision' tests of QCD have become possible. These tests include
measurements of $\as$ and its energy dependence, detailed analyses of two-jet
and three-jet events and studies of QCD coherence.

Further improvements in the field require mainly progress in the theory. Future
${\cal O}(\as^3)$ calculations for event shapes and jet cross sections can be
very useful to check the consistency of the present NLO analyses
[\ref{Zurich}].
Accurate studies of multijet ($n \geq 4$) final states can be performed only
if the corresponding QCD predictions in NLO become available. The increasing
capabilities of tagging $b$-quark jets by micro-vertex detectors allow to
carry out precise experimental studies in this field, but complete NLO
calculations are still missing [\ref{Maina}].

This list does not exhaust the demands to the theory. One of the main problems
indeed remains our poor understanding of the hadronization process and, in
general, of the non-perturbative region. In order to deal with this problem we
need better theoretical methods and more phenomenological investigations. In
this respect, studies of infrared and collinear safe observables at the
boundary
of the perturbative region can be of some help. As briefly discussed at the end
of Sect.~6, the small-$y_0$ behaviour of the subjet multiplicity in two-jet
and three-jet events is an example of this kind of studies. Further
calculations
and predictions for observables which are suitable for similar investigations
are warranted.

\vskip 4.0mm
{\bf \noindent Acknowledgements}
\vskip 2.0mm

I wish to thank Yu.L.\ Dokshitzer, F.\ Fiorani, M.\ Olsson,
L.\ Trentadue, G.\ Turnock and B.R.\ Webber for a long collaboration on
jet physics. I am also grateful to S.\ Banerjee, S.\ Bethke, P.N.\ Burrows,
G.\ Cowan, T.\ Hebbeker, R.\ Miquel, M.\ Schmelling, D.R.\ Ward and
A.\ Wehr for many useful discussions. This research is supported in part
by the EC Programme ``Human Capital and Mobility", Network
``Physics at High Energy Colliders", contract CHRX-CT93-0357 (DG 12 COMA).

\vskip 4.0mm
{\bf \noindent References}
\vskip 2.0mm

\begin{enumerate}

\item \label{Heb}
T.\ Hebbeker, \prep{217}{69}{92}.

\item \label{Gat}
S.\ Bethke, in Proc. of the Aachen Conf. {\it QCD - 20 Years Later}, eds. P.M.\
Zerwas and H.A.\ Kastrup (World Scientific, Singapore, 1993), pag.~43;
%B.R.\ Webber,
%in Proc. of the Aachen Conf. {\it QCD - 20 Years Later}, eds. P.M.\
%Zerwas and H.A.\ Kastrup (World Scientific, Singapore, 1993), pag.~73.
preprint PITHA 94/29.

\item \label{Drees}
J.\ Drees, in Proc. of the Aachen Conf. {\it QCD - 20 Years Later}, eds. P.M.\
Zerwas and H.A.\ Kastrup (World Scientific, Singapore, 1993), pag.~106.

\item \label{Webber}
B.R.\ Webber, plenary talk at the 27th Int. Conf. on High Energy Physics,
Glasgow, July 1994 (quoted as ICHEP 94 in the following),
preprint Cavendish--HEP--94/15.
%\cav{94/15}.

\item \label{SW}
G.\ Sterman and S.\ Weinberg, \prl{39}{1436}{77}.

\item \label{BC}
S.\ Bethke and S.\ Catani, in {\it Perturbative QCD and Hadronic Interactions},
Proc. 27th Rencontres de Moriond, ed. J. Tran Thanh Van (Editions Frontieres,
Gif-sur-Yvette, 1992), pag.~203.
%; I.\ Hinchliffe, preprint LBL-33952, to appear in
%{\it Electroweak Interactions
%and Unified Theories}, Proc. 28th Rencontres de Moriond, Les Arcs, March 1993.

\item \label{Alt}
G.\ Altarelli, in Proc. of the Aachen Conf. {\it QCD - 20 Years Later}, eds.
P.M.\ Zerwas and H.A.\ Kastrup (World Scientific, Singapore, 1993), pag.~172.

\item \label{Bethke}
S.\ Bethke, in Proc. of the 26th Int. Conf. on High Energy Physics, Dallas,
August 1992, ed. J.R.\ Sanford (AIP, New York, 1993), pag.~81.

\item \label{EPS}
S.\ Catani, in Proc. of the Int. Europhysics Conf. on High Energy Physics,
HEP 93, Marseille, July 1993, eds. J.\ Carr and M.\ Perrottet
(Editions Frontieres, Gif-sur-Yvette, 1994), pag.~771.

\item \label{PDG}
Particle Data Group, Review of Particle Properties, \pr{50}{1297}{94}.

\item \label{Mont}
S.\ Bethke, talk at the Conference {\it QCD 94}, Montpellier, July 1994,
preprint PITHA 94/30.

\item \label{OPE}
K.\ Wilson, Phys. Rev. 179 (1969) 1499; K.\ Symanzik, Comm. Math. Phys. 32
(1971) 49; C.\ Callan, \pr{5}{3302}{1972}; R.\ Brandt, Fortschr. Phys. 18
(1970) 249;
M.A.\ Shifman, A.I.\ Vainshtein and V.I.\ Zakharov, \np{147}{385}{79}, 448,
519.

\item \label{MC}
T.\ Sj\"{o}strand, \cpc{39}{347}{86}, \cpc{43}{367}{87}; M.\ Bengtsson and
T.\ Sj\"{o}strand, \np{289}{810}{87}.

\item \label{MC2}
B.R.\ Webber, \np{238}{492}{84};
G.\ Marchesini and B.R.\ Webber, \np{310}{461}{88};
G.\ Marchesini et al., \cpc{67}{465}{92}.

\item \label{MC3}
U.\ Pettersson, preprint LU TP 88-5;
L.\ L\"{o}nnblad and U.\ Pettersson, preprint LU TP 88-15;
L.\ L\"{o}nnblad, preprint LU TP 89-10.

\item \label{MC4}
R.\ Odorico, \cpc{32}{139}{84}, \cpc{59}{527}{90};
P.\ Mazzanti and R.\ Odorico, \np{370}{23}{92}.

\item \label{Pas}
T.\ Hebbeker, M.\ Martinez, G.\ Passarino and G.\ Quast,
\pl{331}{165}{94}.

\item \label{singlet}
K.G.\ Chetyrkin and O.V.\ Tarasov, \pl{327}{114}{94};
S.A.\ Larin, T.\ van Ritbergen and J.A.M.\ Vermaseren,
\pl{320}{159}{94}.

\item \label{Gor}
S.G.\ Gorishny, A.L.\ Kataev and S.A.\ Larin, \pl{259}{144}{91}; L.R.\
Surguladze and M.A.\ Samuel, \prl{66}{560}{91}.

\item \label{Kuhn}
B.A.\ Kniel and J.H.\ K\"uhn, \pl{224}{229}{89}, \np{329}{557}{90}.

\item \label{mass1}
K.G.\ Chetyrkin and A.\ Kwiatkowski, \pl{305}{285}{93};
K.G.\ Chetyrkin, \pl{307}{169}{93}; K.G.\ Chetyrkin and J.H.\ K\"uhn,
\pl{308}{127}{93}; K.G.\ Chetyrkin and A.\ Kwiatkowski, \pl{319}{307}{93}.

\item \label{mass2}
K.G.\ Chetyrkin and J.H.\ K\"uhn, preprint TTP 94-08.

\item \label{BHM}
G.\ Burgers, W.\ Hollik and M.\ Martinez, program BHM; W.\ Hollik,
Fortschr. Phys. 38 (1980) 3, 165; D.\ Bardin et al., program ZFITTER,
version 4.6, preprint CERN-TH 6443/92; G. Montagna et al., program
TOPAZ0, Phys. Commun. 76 (1993) 328.

\item \label{Schaile}
D.\ Schaile, plenary talk at ICHEP 94.
%the 27th Int. Conf. on High Energy Physics, Glasgow, July 1994.

\item \label{BNP}
E.\ Braaten, S.\ Narison and A.\ Pich, \np{373}{581}{92};
F.\ Le Diberder and A.\ Pich, \pl{289}{165}{92}.

\item \label{ALEPH}
ALEPH Coll., D.\ Buskulic et al., \pl{307}{209}{93}.

\item \label{fer}
E.\ Fern\'andez, in Proc. of the Int. Europhysics Conf. on High Energy Physics,
HEP 93, Marseille, July 1993, eds. J.\ Carr and M.\ Perrottet
(Editions Frontieres, Gif-sur-Yvette, 1994), pag.~705.

\item \label{Narison}
S.\ Narison, preprint CERN-TH.7188/94; A.\ Pich,
talk at the Conference {\it QCD 94}, Montpellier, July 1994.

\item \label{Truong}
T.N.\ Truong, preprint EP-CPTh.A266-1093 (EP-CPTh.A266-1093-REV).

\item \label{Alt2}
G.\ Altarelli, preprint CERN-TH.7246/94.

\item  \label{KN}
Z.\ Kunszt, P.\ Nason, G.\ Marchesini and B.R.\ Webber, in
`Z Physics at LEP 1', CERN 89-08, vol.~1, p.~373.

\item  \label{Webber2}
B.R.\ Webber, preprint Cavendish--HEP--94/7.

\item \label{EEC}
A.\ Ali and F.\ Barreiro, \pl{118}{155}{82}, \np{236}{269}{84};
D.G.\ Richards, W.J.\ Stirling and S.D.\ Ellis, \pl{119}{193}{82},
\np{229}{317}{83}; N.K.\ Falk and G.\ Kramer, \zp{42}{459}{89}.

\item \label{Glov}
E.W.N.\ Glover and M.R.\ Sutton, preprint DTP/94/80.

\item \label{ERT}
R.K.\ Ellis, D.A.\ Ross and A.E.\ Terrano,
\np{178}{421}{81}.

\item \label{fo}
L3 Coll., B. Adeva et al., \pl{248}{464}{90}, \pl{257}{469}{91};
ALEPH Coll., D.\ Decamp et al., \pl{255}{623}{91},
\pl{257}{479}{91};
DELPHI Coll., P.\ Abreu et al., \zp{54}{55}{92}; OPAL Coll.,
P.D.\ Acton et al., \zp{55}{1}{92}; S.\ Bethke and J.E.\ Pilcher, Ann. Rev.
Nucl. Part. Sci. 42 (1992) 251.

\item \label{fo2}
Mark-II Coll., S.\ Komamiya et al., \prl{64}{987}{90};
DELPHI Coll., P.\ Abreu et al., \pl{311}{408}{93};
OPAL Coll., R.\ Akers et al., \zp{63}{197}{94}.
%\item \label{fo3}
%
%\item \label{fo4}

\item \label{SLD}
SLD Coll., K.\ Abe et al., \prl{71}{2528}{93}, preprints SLAC-Pub-6451 (1994),
SLAC-Pub-6641 (1994).

\item \label{Bin}
P.\ Binetruy, \pl{91}{245}{80}.

\item \label{thrust}
S.\ Catani, G.\ Turnock, B.R.\ Webber and L.\ Trenta\-due, \pl{263}{491}{91}.

\item \label{hjm}
S.\ Catani, G.\ Turnock and B.R.\ Webber, \pl{272}{368}{91}, \pl{295}{269}{92}.

\item \label{Fiore}
J.\ Kodaira and L.\ Trentadue, \pl{123}{335}{82}; J.C.\ Collins and
D.E.\ Soper, \np{197}{446}{82}, \np{284}{253}{87}; G.\ Turnock, preprint
Cavendish--HEP--92/3;
R.\ Fiore, A.\ Quartarolo and L.\ Trenta\-due, \pl{294}{431}{92}.

\item \label{CDOTW}
S.\ Catani, Yu.L.\ Dokshitzer, M.\ Olsson, G.\ Tur\-nock and B.R.\ Webber,
\pl{269}{432}{91}.

\item \label{CDFW}
S.\ Catani, Yu.L.\ Dokshitzer, F.\ Fiorani and B.R.\ Webber,
\np{377}{445}{92}.

\item \label{CTTW}
S.\ Catani, G.\ Turnock, B.R.\ Webber and L.\ Trenta\-due, \np{407}{3}{93}.

\item \label{LEPres}
ALEPH Coll., D.\ Decamp et al., \pl{284}{163}{92};
L3 Coll., O.\ Adriani et al., \pl{284}{471}{92};
OPAL Coll., P.D.\ Acton et al., \zp{59}{1}{93};
DELPHI Coll., P.\ Abreu et al., \zp{59}{21}{93}.

\item \label{Confres}
L3 Coll.,
%S.\ Banerjee and S.\ M\"uller, L3 note 1441, contributed paper
%to the Int. Europhysics Conf. on High Energy Physics,
%HEP 93, Marseille, July 1993;
S.\ Banerjee, in Proc. of the Int. Europhysics Conf. on High Energy Physics,
HEP 93, Marseille, July 1993, eds. J.\ Carr and M.\ Perrottet
(Editions Frontieres, Gif-sur-Yvette, 1994), pag.~299;
%OPAL Coll., Physics note OPAL-PN102, contributed paper
%to the Int. Europhysics Conf. on High Energy Physics,
%HEP 93, Marseille, July 1993;
TOPAZ Coll., Y.\ Ohnishi et al., \pl{313}{475}{93};
TPC/Two-Gamma Coll., D.A.\ Bauer et al., preprint SLAC-Pub-6518 (1994);
TOPAZ Coll., M.\ Aoki et al., contributed paper GLS0944 to ICHEP 94.
%to the 27th Int. Conf. on High Energy Physics, Glasgow, July 1994;

\item \label{JADE}
JADE Coll., W.\ Bartel et al., \zp{33}{23}{86}.

\item \label{BKSS}
S.\ Bethke, Z.\ Kunszt, D.E.\ Soper and W.J.\ Stirling, \np{370}{310}{92};
N.\ Brown and W.J.\ Stirling, \zp{53}{629}{92}.

\item \label{jettop}
S.\ Catani, in {\it QCD at 200 TeV}, Proc. 17th Eloisa\-tron Project Workshop,
eds. L.\ Cifarelli and Yu.L.\ Dokshitzer (Plenum Press, New York, 1992),
pag.~21.

\item \label{DO}
Yu.L.\ Dokshitzer and M.\ Olsson, \np{396}{137}{93}.

\item \label{DURHAM}
{\it Report of the Hard QCD Working Group}, in Proc. Durham Workshop on Jet
Studies at LEP and HERA, \jp{17}{1537}{91}.

\item \label{BS}
N.\ Brown and W.J.\ Stirling, Phys. Lett. 252B (1990) 657.
%\pl{252}{657}{90}.

\item \label{CDW}
S.\ Catani, Yu.L.\ Dokshitzer and B.R.\ Webber, \pl{285}{291}{92}.

\item \label{kthad}
S.\ Catani, Yu.L.\ Dokshitzer, M.H.\ Seymour and B.R.\ Webber,
\np{406}{187}{93}; S.D.\ Ellis and D.E.\ Soper, \pr{48}{3160}{93}.
%preprint CERN-TH.6860/93.

\item \label{Mor}
M.H.\ Seymour, in {\it QCD and High Energy Hadronic Interactions},
Proc. 28th Rencontres de Moriond, ed. J. Tran Thanh Van (Editions Frontieres,
Gif-sur-Yvette, 1993), pag.~141;
S.\ Catani, in {\it QCD and High Energy Hadronic Interactions},
Proc. 28th Rencontres de Moriond, ed. J. Tran Thanh Van (Editions Frontieres,
Gif-sur-Yvette, 1993), pag.~257;
B.R.\ Webber, \jp{19}{1567}{93}.

\item \label{Sey}
M.H.\ Seymour, \zp{62}{127}{94}, \np{421}{545}{94}.

\item \label{SA}
J.\ Huth et al., in Proc. of {\it Research Directions for the Decade, Snowmass
1990}, ed. E.L.\ Berger (World Scientific, Singapore, 1992), pag.~134.

\item \label{colour}
OPAL Coll., M.Z.\ Akrawy et al., \zp{49}{49}{90};
L3 Coll., B.\ Adeva et al., \pl{248}{227}{90};
ALEPH Coll., D.\ Decamp et al., \pl{284}{151}{92};
DELPHI Coll., P.\ Abreu et al., \zp{59}{357}{93},
contributed paper GLS0180 to ICHEP 94;
OPAL Coll., contributed paper GLS0597 to ICHEP 94.

\item \label{Kra}
G.\ Kramer, {\it Springer Tracts in Modern Physics} (1984) 102.

\item \label{qgjet}
OPAL Coll., P.D.\ Acton et al., \zp{58}{387}{93};
DELPHI Coll., K.\ Hamacher et al., contributed paper GLS0187 to ICHEP 94;
ALEPH Coll., contributed paper GLS0539 to ICHEP 94;
SLD Coll., Y.\ Iwasaki and U.\ Tohoku, SLAC-Pub-6597 (1994).

\item \label{ALEPHg}
ALEPH Coll., contributed paper GLS0545 to ICHEP 94.

\item \label{Mue}
A.H.\ Mueller,  \np{213}{85}{83} (Erratum \np{241}{141}{84}),
\np{228}{351}{83)}; Yu.L.\ Dokshitzer, V.A.\ Khoze and S.I.\ Troyan, in
{\it Perturbative Quantum Chromodynamics}, ed. A.H.\ Mueller (World Scientific,
Singapore, 1989), pag.~241.

\item \label{sub}
S.\ Catani, B.R.\ Webber, Yu.L.\ Dokshitzer and F.\ Fiorani, \np{383}{419}{92}.

\item \label{23jet}
L3 Coll., S.\ Banerjee, L3 Note 1171 contributed paper to
the 26th Int. Conf. on High Energy Physics, Dallas, August 1992;
ALEPH Coll., D.\ Decamp et al., Note A845 contributed paper to
the 26th Int. Conf. on High Energy Physics, Dallas, August 1992;
L3 Coll., O.\ Adriani et al., \prep{236}{1}{93};
OPAL Coll., R.\ Akers et al., preprint CERN-PPE/94-52.

\item \label{string1}
JADE Coll., W.\ Bartel et al., \pl{101}{129}{81}.

\item \label{string2}
B.\ Andersson, G.\ Gustafson and T.\ Sj\"{o}strand, \pl{94}{211}{80},
\zp{6}{235}{80}; Ya.I.\ Azimov, Yu.L.\ Dokshitzer, V.A.\ Khoze and S.I.\
Troyan,
\pl{165}{147}{85}.

\item \label{string3}
DELPHI Coll., P.\ Aarnio et al., \pl{240}{271}{90};
OPAL Coll., M.Z.\ Akrawy et al., \pl{261}{334}{91};
L3 Coll., B.\ Adeva et al., \zp{55}{39}{92}.

\item \label{Zurich}
{\it New Techniques for Calculating Higher Order QCD Corrections}, Proc. ETH
Workshop, Z\"urich 1992, ed. Z.\ Kunszt, preprint ETH-TH/93-01.

\item \label{Maina}
A.\ Ballestrero, E.\ Maina and S.\ Moretti, \pl{294}{425}{92}.

\end{enumerate}

\end{document}